\documentclass[12pt]{article}
\usepackage{graphicx}
\usepackage{xcolor}
\usepackage{hyperref}

\newcommand\pubnumber{CIPANP2015-Levillain}
\newcommand\pubdate{\today}

\def\Title#1{\begin{center} {\Large #1 } \end{center}}
\def\Author#1{\begin{center}{ \sc #1} \end{center}}
\def\Address#1{\begin{center}{ \it #1} \end{center}}

\newcommand\pubblock{\rightline{\begin{tabular}{l} \pubnumber\\
         \pubdate  \end{tabular}}}
\newenvironment{Abstract}{\begin{quotation}  }{\end{quotation}}
\newenvironment{Presented}{\begin{quotation} \begin{center} 
             PRESENTED AT\end{center}\bigskip 
      \begin{center}\begin{large}}{\end{large}\end{center} \end{quotation}}

\begin{document}
\begin{titlepage}
\pubblock

\vfill
\Title{$A_{LL}(p_T)$ for single hadron photoproduction at high $p_T$}
\vfill
\Author{Maxime Levillain \footnote{On behalf of the COMPASS Collaboration}}
\Address{CEA Saclay, DSM/IRFU/SPhN/LSN F-91191 Gif sur Yvette, FRANCE}
\vfill
\begin{Abstract}
In order to understand the gluon contribution to the nucleon spin, some experiments can study the production of hadrons at high transverse momemtum from lepton-nucleon or nucleon-nucleon scattering. RHIC has recently measured such double spin asymmetries $A_{LL}(p_T)$ for pion production at high center of mass energies \cite{star} \cite{phenix}, and inclusion of its data to global fits based on NLO collinear pQCD calculations gives some constraints on the gluon polarization in the range $0.05<x_g<0.2$. \cite{dssv++}\\

To complement these results at COMPASS range, we will present preliminary COMPASS results on double longitudinal spin asymmetries $A_{LL}(p_T)$ for single hadron production measured on deuteron and proton target at $Q^2<1$~GeV$^2$, $p_T>1$~GeV/c and center of mass energy $\sqrt{s}\approx 18$~GeV. All COMPASS data taken from 2002 to 2011 by scattering 160 GeV polarized muons on longitudinally polarized $^6$LiD and NH$_3$ targets have been used, and the number of hadrons collected with $p_T>1$~GeV/c for this analysis amounts to about 10 millions. The obtained asymmetries will be compared to theoretical predictions of at NLO without gluon resummation calculation.
\end{Abstract}
\vfill
\begin{Presented}
CIPANP2015\\
Vail, Colorado, USA,  May 19--24, 2015
\end{Presented}
\vfill
\end{titlepage}
\def\thefootnote{\fnsymbol{footnote}}
\setcounter{footnote}{0}

\section*{Introduction}

The data used in this analysis come from the COMPASS experiment, located at CERN. It uses a polarized muon beam coming from the SPS at 160 GeV (2002-2007) or 200 GeV (2011) scattered off a polarized $^6$LiD (2002-2006) or NH$_3$ (2007-2011) target. The COMPASS spectrometer allows us to select precisely the data at $p_T>1$~GeV/c and $Q^2<1$~GeV/c$^2$ with also other kinematical and geometrical cuts to place oneself in the theoretical settings and have cleaner data. Among these cuts, the main ones are on $y$ the fractional energy loss ($0.1<y<0.9$) to avoid large radiative contributions and region with imperfect Monte Carlo; on $z$ the fractional hadron energy ($0.2<z<0.8$) to avoid current fragmentation and exclusive events; and finally on $\theta_h$ ($0.01<\theta_h<0.12$~rad) which is the only region well covered by the spectrometer (especially for 2002-2004 data). \\     

The previous studies of these data were using the Monte Carlo generator PYTHIA to unfold the partonic information at LO from $A_{LL}$. This new study allows to extract these information at NLO thanks to a new theoretical framework that will be followingly presented. 

\section{Theoretical Framework}

\begin{figure}[htb]
 \begin{minipage}[b]{0.5\linewidth}
  \centering
  \includegraphics[width=4cm]{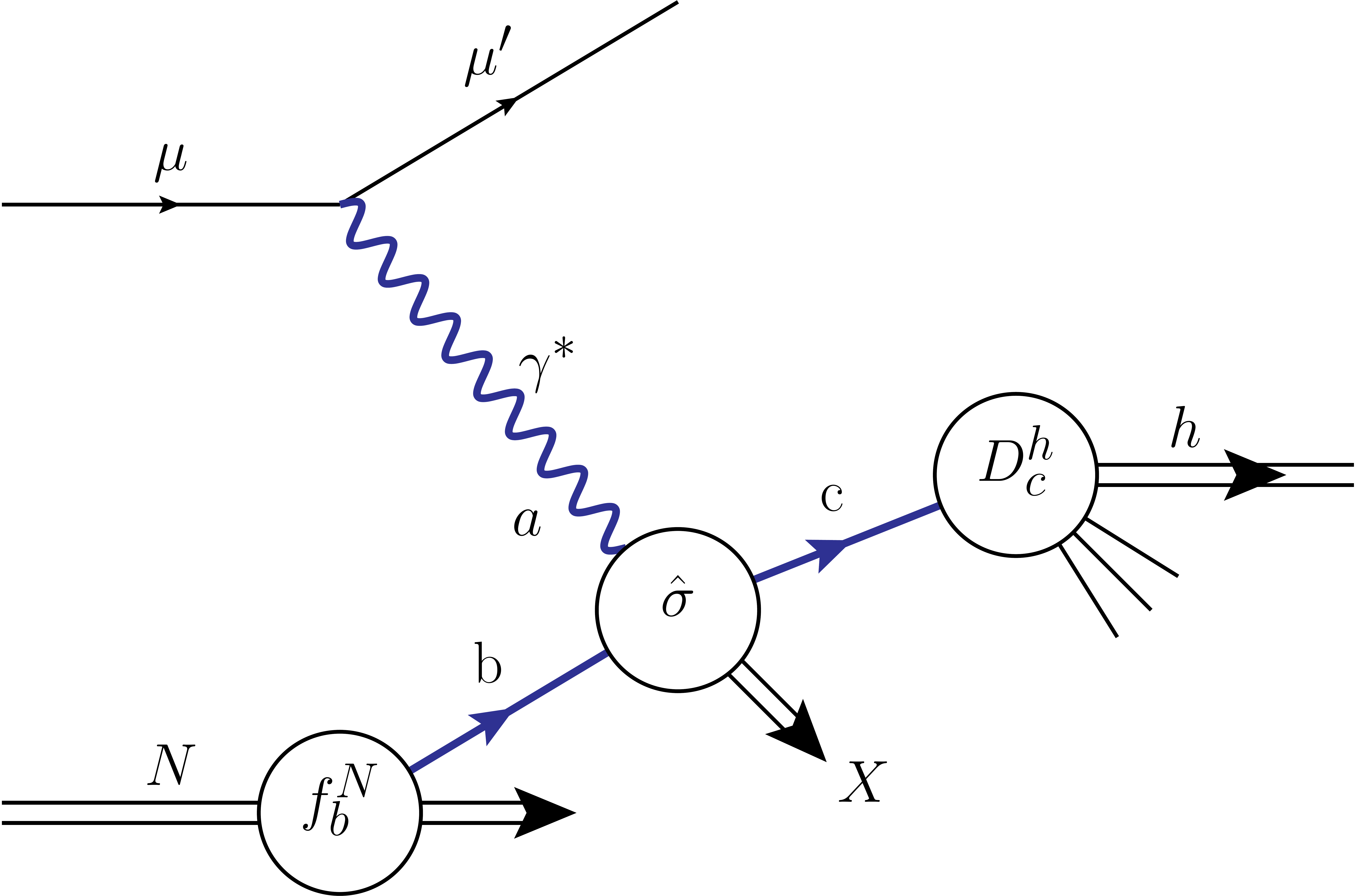}
  \caption{direct $\gamma - contribution$ \label{direct}}
 \end{minipage}
 \begin{minipage}[b]{0.5\linewidth}
  \centering
  \includegraphics[width=4cm]{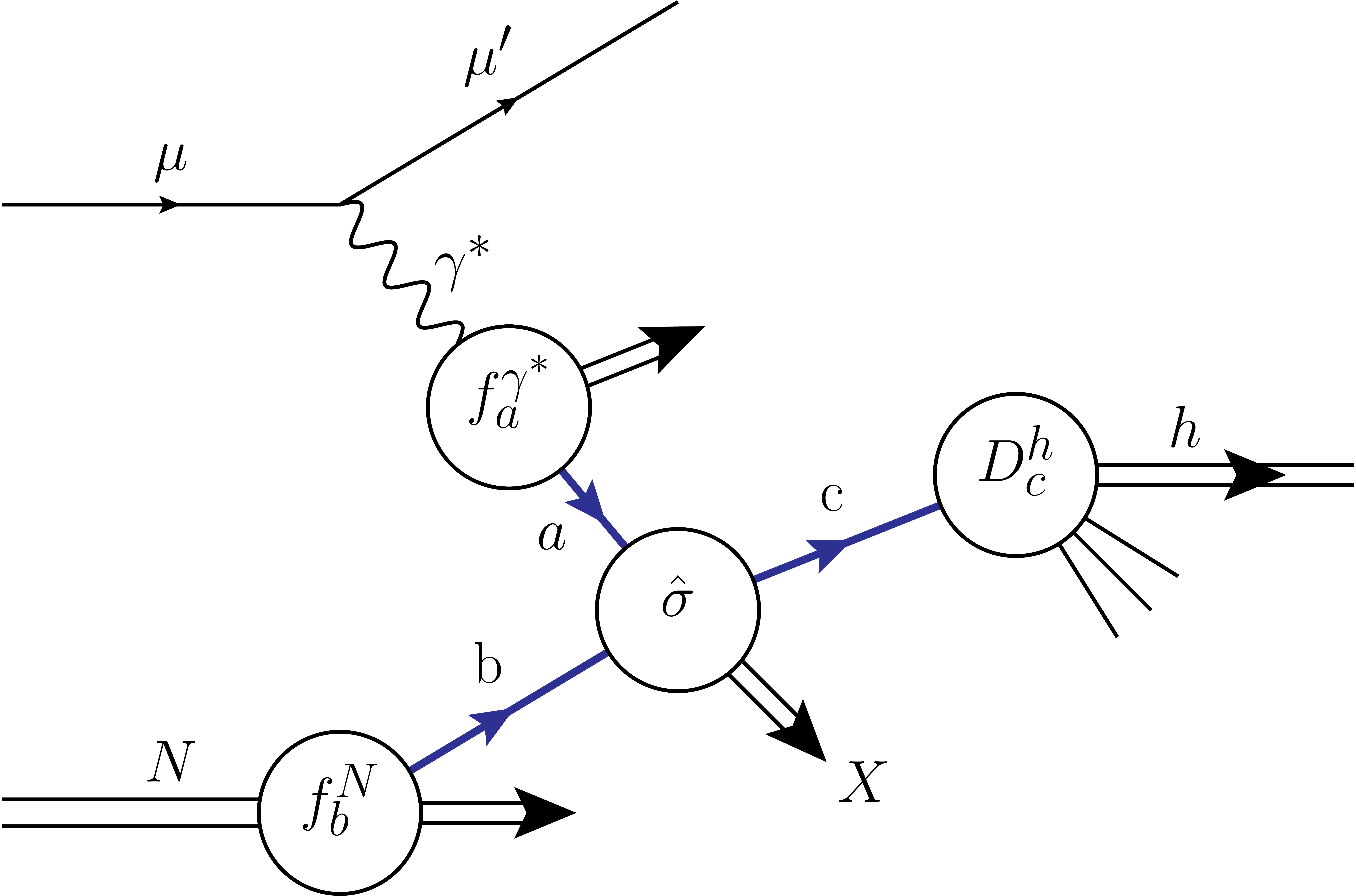}
  \caption{resolved $\gamma - contribution$ \label{resolved}}
 \end{minipage}
\end{figure}

This analysis is based on collinear perturbative QCD calculations done by B.~J\"ager, M.~Stratmann and W.~Vogelsang in 2005 \cite{vogelsang}. Unpolarized and polarized cross-sections can be computed in function of $p_T$ and $\eta$ (the pseudo-rapidity) by convoluting nucleon PDFs, with perturbative partonic cross-section, fragmentation function distribution, and  a last part $\Delta f_a^{\mu}$ corresponding to a '\textit{muon distribution function}'.
This last distribution leads to two different parts of the calculations: 
\begin{itemize}
\item Direct $\gamma - contribution$ illustrated in Fig.~\ref{direct} with $\Delta f_a^{\mu}(x_a,\mu_f) = \int_{x_a}^1 \frac{dy}{y} \Delta P_{\gamma \mu}(y)\delta (1-x_{\gamma})$, convolution of Weis\"acker-Williams probability with a Dirac distribution.
\item Resolved $\gamma - contribution$ illustrated in Fig.~\ref{resolved} with $\Delta f_a^{\mu}(x_a,\mu_f) = \int_{x_a}^1 \frac{dy}{y} \Delta P_{\gamma \mu}(y) \Delta f_a^{\gamma}(x_{\gamma}=\frac{x_a}{y},\mu_f)$: convolution of Weis\"acker-Williams probability with a perturbative photon-parton distribution.
\end{itemize}
\begin{equation}
 \frac{d\Delta \sigma^h}{d\sigma^h}(p_T,\eta) =  \frac{\sum _{a,b,c} \textcolor{red}{\Delta f_a^{\mu}} \otimes \textcolor{red}{\Delta f_b^N} \otimes d\Delta \hat{\sigma}_{a,b\rightarrow c,X} \otimes \textcolor{red}{D_c^h}}{\sum _{a,b,c} \textcolor{red}{f_a^{\mu}} \otimes \textcolor{red}{f_b^N} \otimes d\hat{\sigma}_{a,b\rightarrow c,X} \otimes \textcolor{red}{D_c^h}}  = \frac{d\Delta \sigma_{dir} + d \Delta \sigma_{res}}{d\sigma_{dir}+d\sigma_{res}}
\end{equation}

The collinear assumption implies that the measured $p_T$ should arise from hard subprocesses which can then only be higher order ones. The leading order $\alpha^2\alpha_S$ includes 2 direct processes: the photon-gluon fusion ($\gamma^* g \rightarrow q\bar{q}$) which probes the gluon distribution and the QCD Compton ($\gamma^* q \rightarrow qg$), but also other resolved processes taking into account the gluon distribution. The next-to leading order $\alpha \alpha_S^2$ with a multitude of direct and resolved processes, makes appear leading logarithms $\alpha \alpha_S^klog^{2k-1}(w/(1-w))$ which can become large at low center of mass energy and need to be checked out. \\

\begin{figure}[htb]
 \begin{minipage}[b]{0.5\linewidth}
 \centering
 \includegraphics[width=7cm]{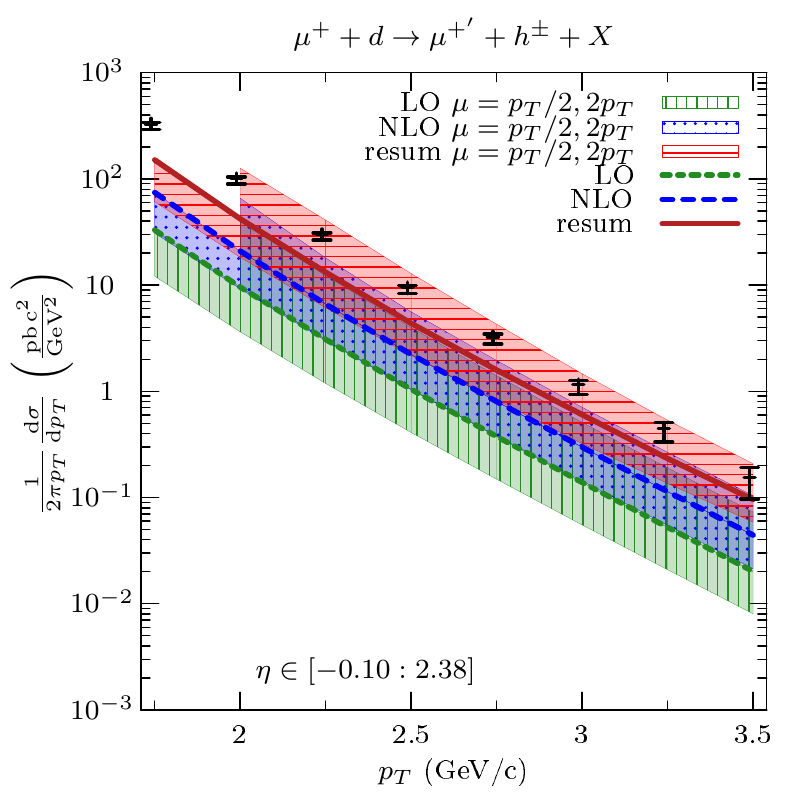}
 \end{minipage}
 \begin{minipage}[b]{0.5\linewidth}
 \centering
 \includegraphics[width=7cm]{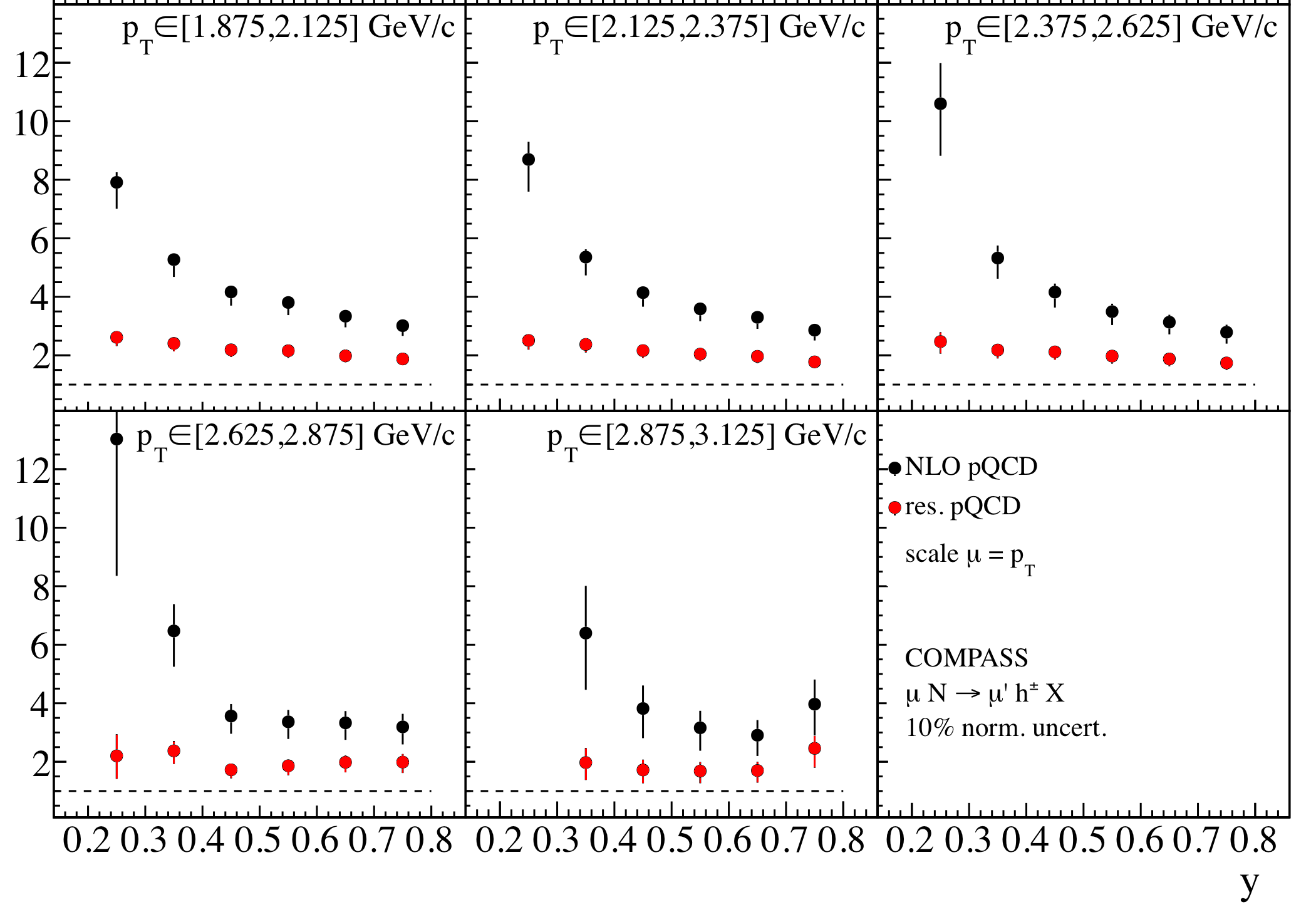}
 \end{minipage}
 \caption{Unpolarized cross-section measurements compared with theoretical calculations at LO, NLO and NLO+NLL (left), and ratios experiment/theory for NLO and NLO+NLL (right) \label{unpol}}
\end{figure}

For this purpose a preliminary study has been done on the unpolarized cross-section \cite{hoeppner} to check if this theory used in RHIC ($\sqrt{s} \approx 200\,GeV$) was still applicable in COMPASS kinematics ($\sqrt{s} \approx 18\,GeV$). The study concludes that in COMPASS kinematics, gluon resummation is needed to explain the cross-section and its $y-dependency$ (Fig.~\ref{unpol}) \cite{pfeuffer}.

\section{Measurement Method}

The asymmetry extracted from COMPASS data is a lepton-nucleon semi-inclusive asymmetry based on hadron counts.
This asymmetry is diluted by \textcolor{red}{$\beta$} composed of 3 factors: $f$ the dilution factor of the target, $P_\mu$ the beam polarization and $P_x$ the target polarization (divided in cells with opposite polarizations). These different cells give us two hadron counts ($N_u$ and $N_d$) and two other after rotation of the solenoid field ($N_{u'}$ and $N_{d'}$).

\begin{eqnarray}
 \nonumber
 N_x  = \phi a_x n_x \sigma_0 (1+(\textcolor{blue}{f \cdot P_{\mu}} \cdot P_x) A_{LL}) & \rightarrow & A_{raw} = \frac{N_u-N_d}{N_u+N_d}
\end{eqnarray}

To remove more efficiently the acceptance effects between the cells, a second order method is used to extract the asymmetry from the quantity $\delta$: 
\begin{equation}
 \nonumber
 \delta = \frac{N_u \cdot N_{d'}}{N_d \cdot N_{u'}} \approx \frac{(1+\langle \textcolor{red}{\beta_u} \rangle A_{LL})(1+\langle \textcolor{red}{\beta_{d'}} \rangle A_{LL})}{(1+\langle \textcolor{red}{\beta_d} \rangle A_{LL})(1+\langle \textcolor{red}{\beta_{u'}} \rangle A_{LL})}
\end{equation}

To statistically optimize $\langle A_{LL} \rangle$ one weights each hadron by \textcolor{blue}{$w = f \cdot P_\mu$} to get $\frac{\sigma_{A_{\textcolor{blue}{w}}}}{\sigma_{A_{st}}} \approx \sqrt \frac{\langle \textcolor{blue}{w} \rangle^2}{\langle \textcolor{blue}{w}^2 \rangle} $:
\begin{equation}
 \nonumber
 \delta = \frac{\sum \textcolor{blue}{w_u} \cdot \sum \textcolor{blue}{w_{d'}} }{\sum \textcolor{blue}{w_d} \cdot \sum \textcolor{blue}{w_{u'}} } \approx \frac{(1+\langle \textcolor{red}{\beta_u} \rangle_{\textcolor{blue}{w}} \cdot A_{LL})(1+\langle \textcolor{red}{\beta_{d'}} \rangle_{\textcolor{blue}{w}}  \cdot A_{LL})}{(1+\langle \textcolor{red}{\beta_d} \rangle_{\textcolor{blue}{w}} \cdot A_{LL})(1+\langle \textcolor{red}{\beta_{u'}} \rangle_{\textcolor{blue}{w}} \cdot A_{LL})}
\end{equation}

Asymmetry is finally extracted for each group of data separated by two solenoid field rotations and then averaged to remove long term instability:
\begin{equation}
 A_{LL} = \frac{\sum_i \sigma_{A_i}^{-2} A_i}{\sum_i \sigma_{A_i}^{-2}}
\end{equation}

\section{Systematic Studies}

A various of qualitative studies have been done on '\textit{false}' asymmetries (non physical asymmetries supposed to be 0):
\begin{itemize}
\item Misconfiguration asymmetries (grouping of data with identical spin state supposed to give a 0 asymmetry) showing the apparatus asymmetry.
\item Difference of asymmetry between left part and right part of the spectrometer, and difference of asymmetry between top part and bottom part of the spectrometer, showing efficiency anisotropy of the spectrometer.
\item Difference between upstream cell and downstream cell showing acceptances effects and polarization inhomogeneity of the target.
\item Difference between day and night asymmetries showing effects of thermal expansion and electronic noise fluctuation.
\end{itemize}
In conclusion, no false asymmetry has been detected for this analysis under the statistical uncertainties.\\
Other quantitative studies have been pursued: 
\begin{itemize}
\item A multiplicative error study: $\Delta A^{mult}_{LL} = A_{LL} \sqrt{\left(\frac{dP_{\mu}}{P_{\mu}}\right)^2+\left(\frac{dP_t}{P_t}\right)^2+\left(\frac{df}{f}\right)^2} \approx\, 0.07\, A_{LL}$ (where $P_t = \langle P_x \rangle_x$).
\item A time stability study through pulls distribution $\Delta r = \frac{A_i - \bar{A}}{\sigma_A^{stat}}$, from which one can extract a systematic error value of the same order as the statistical one for this analysis.
\end{itemize}

\section{Results}

\begin{figure}[htb]
 \centering
 \includegraphics[width=7.5cm]{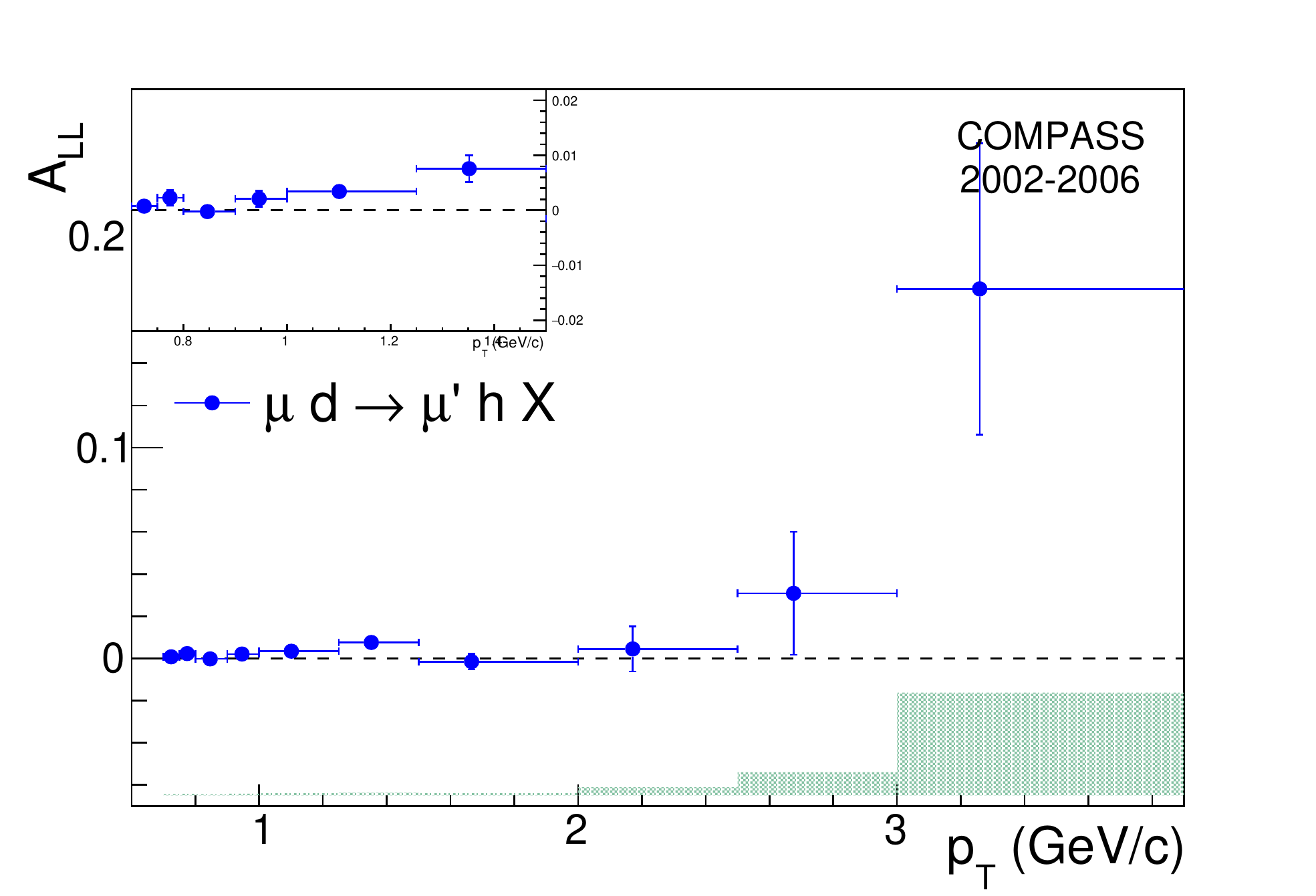}
 \includegraphics[width=7.5cm]{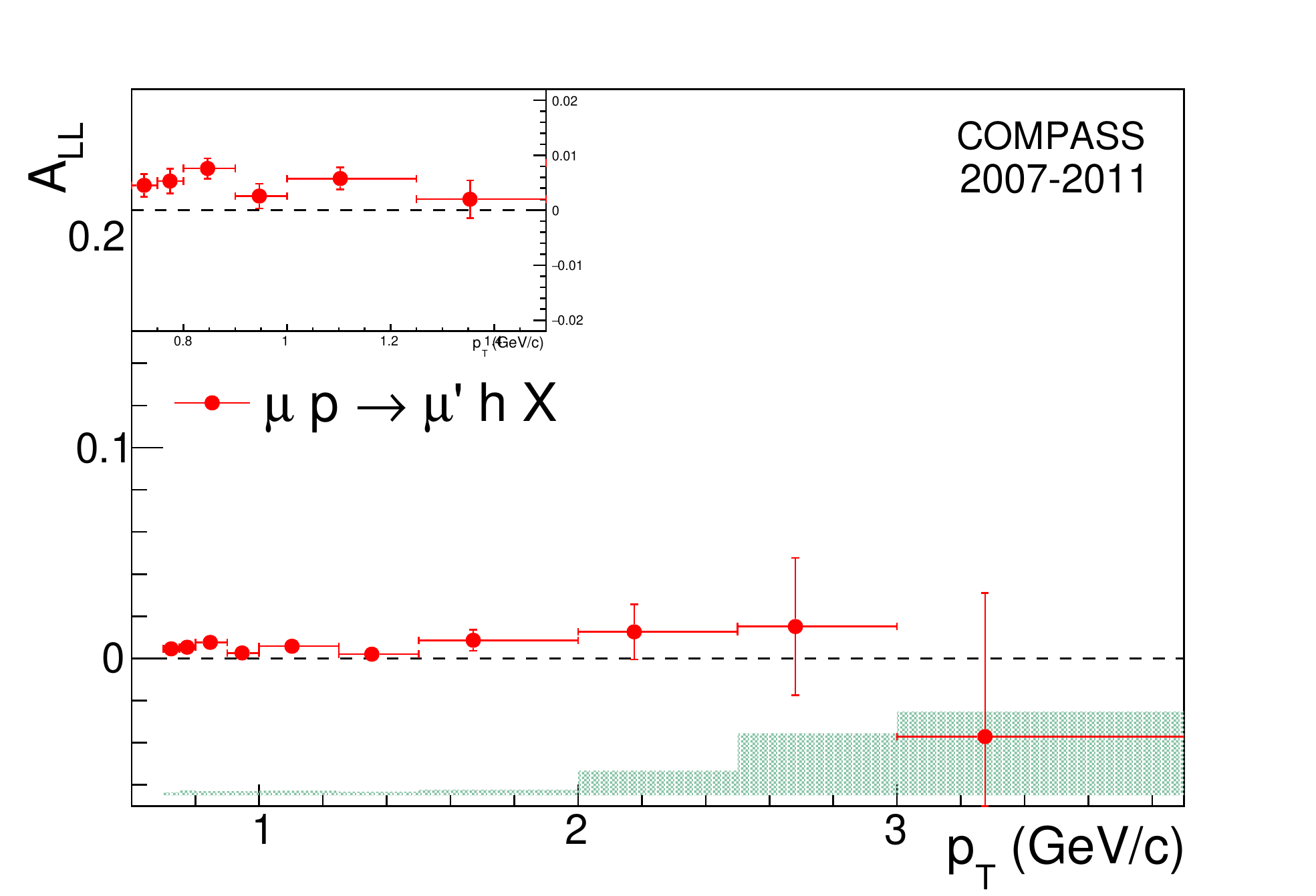}
 \caption{$A_{LL}^d$ for 2002-2006 COMPASS deuteron data (left) and $A_{LL}^p$ for 2007-2011 COMPASS proton data (right) \label{results}}
\end{figure}

Asymmetries have been calculated for each year to find a good compatibility between different setups for deuteron and proton target separately (Fig.~\ref{results}).

These results are on the whole compatible with 0 except for high~$p_T$ deuteron and low~$p_T$ proton. This last observation was in some way expected and observed in other analyses.

\section{Comparison with Theoretical Calculations}

These results have been compared with theoretical calculations using the same code as the one in 2005 with some exceptions:
\begin{itemize}
\item Unpolarized PDFs come from CTEQ6 (as previous calculations).
\item Fragmentation functions distribusions come from DSS.
\item Unpolarized photon-parton distributions come from GRS and polarized ones from W.~Vogelsang code.
\item 3 different sets of polarized PDFs have been used to show the effect of $\Delta G$ on $A_{LL}(p_T)$: 2 extreme sets of GRSV \cite{grsv} and DSSV++ 2014 \cite{dssv++}.
\end{itemize}

These calculations have been of course done to match kinematics of COMPASS, i.e. a Weis\"acker-Williams probability with $Q^2_{max}=1\,(GeV/c)^2$ and cross-sections cut with $0.1<y<0.9$ and $0.2<z<0.8$, and integrated over $\eta_{cms} \in [-0.1,2.4]$ which corresponds to the $\theta_h$ cut of the data.

\begin{figure}[htb]
 \centering
 \includegraphics[width=7.5cm]{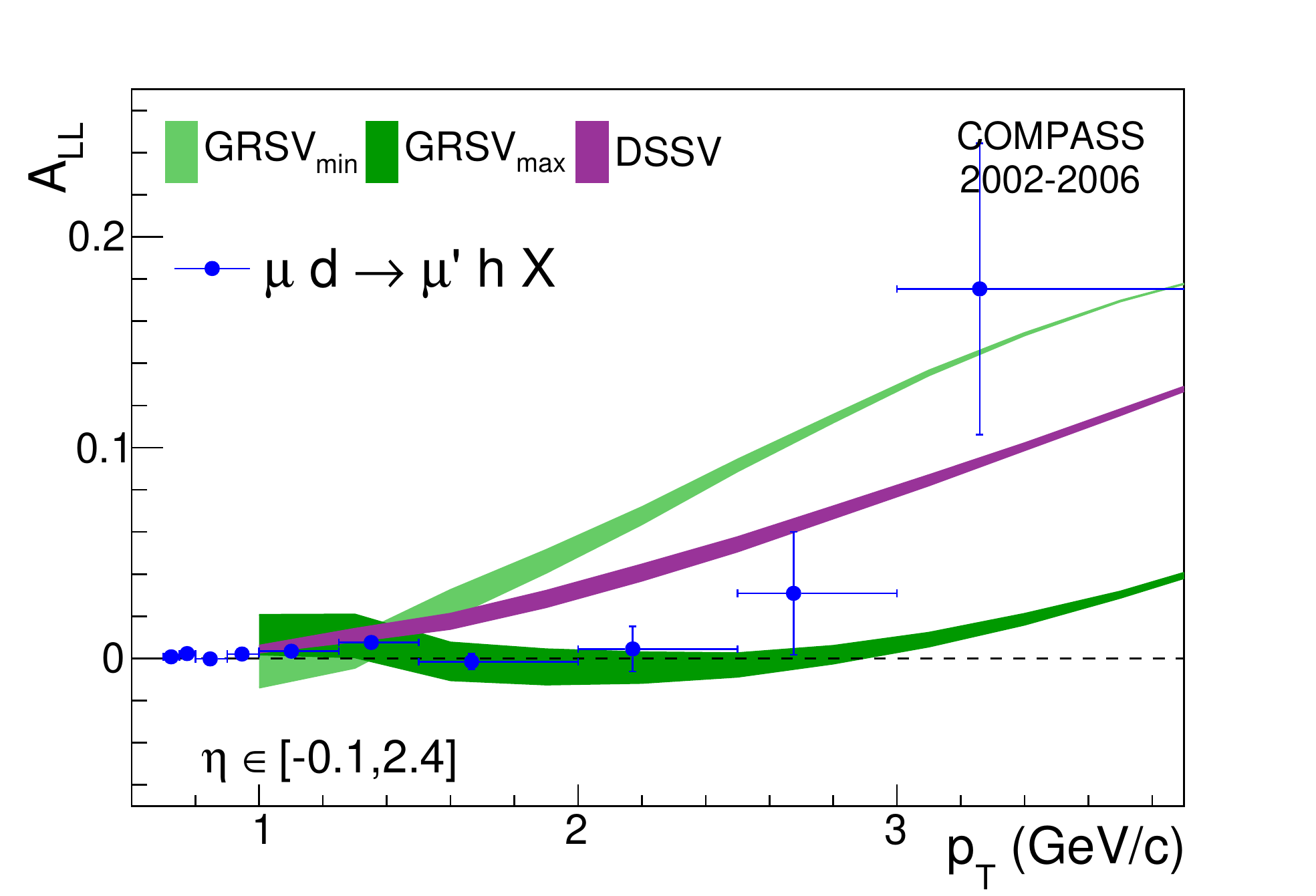}
 \includegraphics[width=7.5cm]{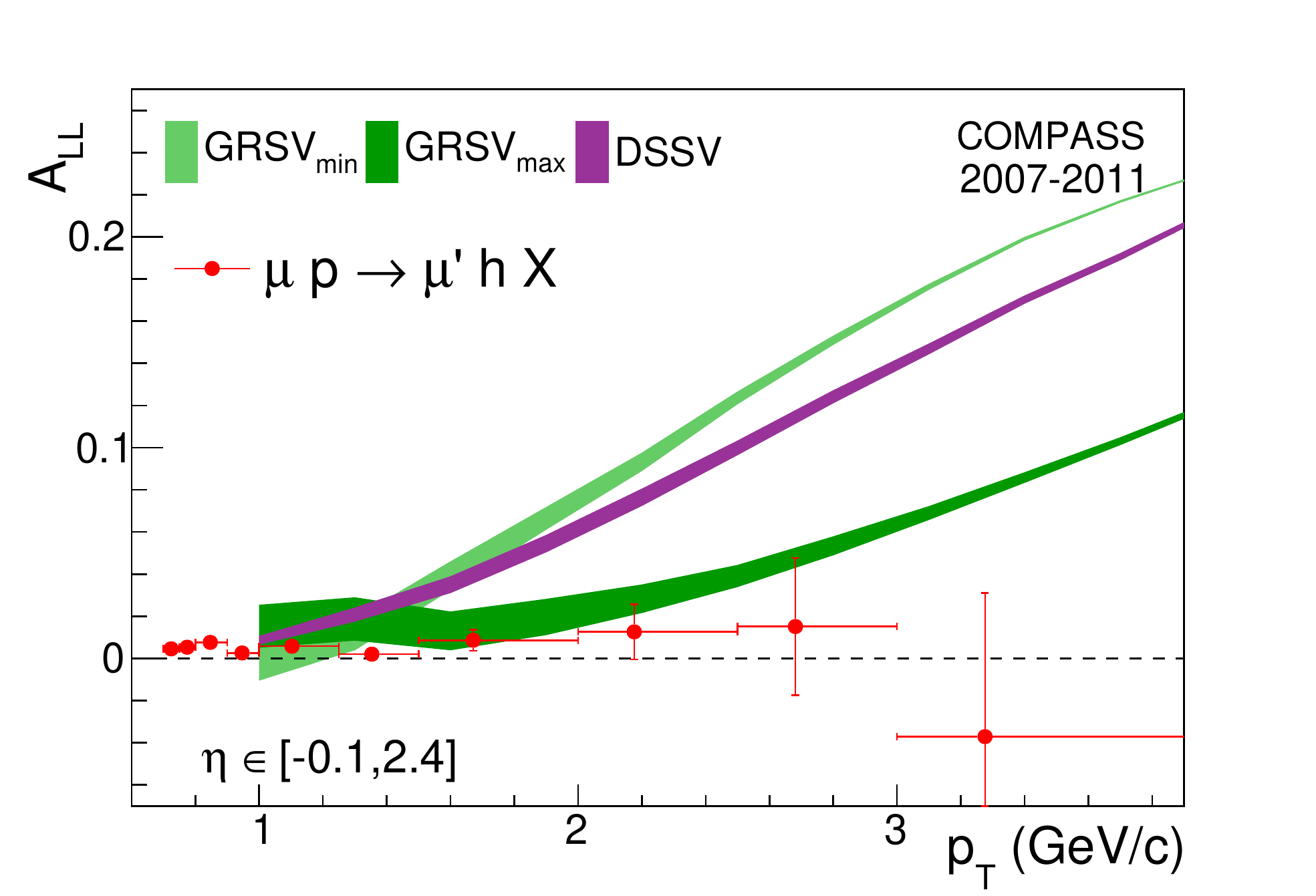}
 \caption{Comparison between experimental $A_{LL}$ and NLO theoretical asymmetries for different sets of polarized PDFs ($A_{LL}^d$ on the left and $A_{LL}^p$ on the right) \label{comp}}
\end{figure}

These first comparisons (Fig.~\ref{comp}) are only preliminary, knowing that the gluon resummation should change the theoretical calculations. However the deuteron data points would indicate a large positive $\Delta G$, whereas the proton data points cannot be explained by any set of PDFs, how large the $\Delta G$ is. This disagreement between the different target material results makes difficult to interpret them.

\section*{Prospects}

Even though this data analysis is already well advanced, some work remains to do to get clearer data by separating charges and flavors. For this purpose it is foreseen to extract from the same data $A_{LL}^{h+}$, $A_{LL}^{h-}$, $A_{LL}^{\pi +}$ and $A_{LL}^{\pi -}$ to compare with the theoretical calculations. Getting $A_{LL}^{k +}$ and $A_{LL}^{k -}$ is also possible but will bring us less information due to the higher statistical uncertainty of these data.

Of course this comparison with theoretical calculations is just a first step to illustrate the link between $A_{LL}$ and $\Delta G$, but these results should be in the future integrated to world data fits in order to reduce uncertainties over $\Delta g(x_g)$  in COMPASS $x_g$ region which is not yet well determined for these data.

Even though this data analysis is already well advanced, some work remains to do to get clearer data by separating charges and flavors. For this purpose it is foreseen to extract from the same data $A_{LL}^{h+}$, $A_{LL}^{h-}$, $A_{LL}^{\pi +}$ and $A_{LL}^{\pi -}$ to compare with the theoretical calculations. Getting $A_{LL}^{k +}$ and $A_{LL}^{k -}$ is also possible but will bring us less information due to the higher statistical uncertainty of these data.

Of course this comparison with theoretical calculations is just a first step to illustrate the link between $A_{LL}$ and $\Delta G$, but these results should be in the future integrated to world data fits in order to reduce uncertainties over $\Delta g(x_g)$  in COMPASS $x_g$ region which is not yet well determined for these data.

\end{document}